\documentclass[12pt]{article}

\def\be{\begin{equation}}
\def\ee{\end{equation}}
\def\bea{\begin{eqnarray}}
\def\eea{\end{eqnarray}}
\def\a{\alpha}
\def\b{\beta}

\def\t{\theta}

\usepackage{amsmath, amsgen,amstext,amsbsy,amsopn,amsthm, amssymb }
\begin{document}

\title{Dimers on two-dimensional lattices}

\author{ F. Y. Wu \\
  Department of Physics\\ Mortheastern University \\
   Boston, 
Massachusetts 02115, USA }

%\date{July 21, 2002}
%\date{\version }
\maketitle
\begin{abstract} \
We consider  
close-packed dimers, or perfect matchings, on two-dimensional regular lattices.  We 
review known results and derive  new expressions for the free energy, entropy,
and the molecular freedom of dimers for a number of lattices including the
simple-quartic $(4^4)$, 
  honeycomb $(6^3)$,  triangular $(3^6)$,  kagom\'e $(3\cdot 6 \cdot 3 \cdot 6)$,
 3-12 $(3\cdot 12^2)$ and its dual $[3\cdot 12^2]$,
and  4-8 $(4\cdot 8^2)$ and its dual Union Jack $[4\cdot 8^2]$
 Archimedean tilings.
The occurrence and nature of phase transitions are also analyzed and discussed.
  
\end{abstract}
 
\newpage
\section{Introduction}
A central problem in statistical physics and combinatorial mathematics 
is the enumeration of close-packed dimers, often referred to as perfect matchings in 
mathematical literature, on lattices
which mimics the adsorption of diatomic molecules on a surface \cite{fr}.
A folklore in lattice statistics
 states that close-packed dimers 
can always be enumerated for  two-dimensional lattices.
Indeed, the seminal works 
of Kasteleyn \cite{kas} and Fisher and Temperley \cite{temp, fisher61}
on the simple-quartic lattice
produced the first exact solution.
 However, a search of the literature indicates that very little else
has been published.   
This paper is an attempt to clarify the situation.  Here we
 review known results and derive new expressions for the free energy, entropy, 
and molecular freedom for various Archimedean two-dimensional lattices.
We also discuss analytic properties of the free energy.
      As we shall see, the task
is not  straightforward as previously thought,
and the  enumeration for many Archimedean lattices still remain open.

 \medskip
We consider a  regular two-dimensional array of $N$ ($=$ even) lattice points
which can be covered by $N/2$ dimers in the large $N$ limit.
Denote the dimer weights by $\{z_i\}$   and define
the generating function 
\be
Z (\{z_i\}) = \sum_{\rm dimer\ coverings} \prod_i z_i^{n_i}
\ee
where the summation is over all dimer coverings and
$n_i$ is the number of dimers with weight $z_i$.
For a large lattice $Z(\{z_i\})$ is expected to grow exponentially in $N$.
Our goal is to compute the  ``free energy" per dimer\be
f(\{z_i\}) = \lim_{N\to\infty} \frac 2 N \ln Z (\{z_i\}). \label{freeE}
\ee
 Setting   $z_i=1$, the numerical value 
\be
S= f (\{1\}) \label{entropy}
\ee
is the entropy of  adsoptions of 
diatomic molecules and
\be 
W= \lim_{N\to\infty} \big[Z_N(\{1\}) \big]^{2/N} = e^S
\ee
is  often known as the per-dimer  molecular freedom.
These are  quantities of primary interest
in mathematics, physics, and chemistry.\footnote{In some earlier papers relevant
quantities are defined on a per lattice site basis with $W_{\rm per\, site} = \sqrt W$,
etc.}

\medskip
The following two integration formulas are useful for our purposes:
 \be
{1\over {2\pi} } \int_0^{2\pi} d\t \ln(2A + 2B\cos \t + 2C \sin \t)
  = \ln \big[ A + \sqrt{A^2 -B^2 -C^2}\big] \label{id1} 
\ee
for real $A,B,C$, and
\be
{1\over {2\pi} } \int_0^{2\pi} d\t \ln \big|A + B e^{i\t} \big| = \ln \max \{ |A|, |B| \}
\label{id2}
\ee
for real or complex $A$ and $B$.

\medskip
We have also the following result which holds generally for any graph, planar or not:
 
\medskip
{\it Proposition:  Let $G$ be a bipartite graph consisting of two sets
of equal number of vertices $A$ and $B$, with vertices in $A$ connected
only to vertices in $B$, and vice versa.
   Let $G^*$ be a graph generated
from $G$ by adding edges connecting vertices within one set.
Let $Z_G$ and $Z_{G^*}$ be the respective dimer
generating functions.
  Then we have the identity
\be
Z_{G^*} = Z_G . \label{prop}
\ee
Namely, the addition of edges connecting  vertices within one set
of vertices in a bipartite graph
 does not alter the dimer generating function.}
 
\medskip
{\it Proof}:
Let the inserted edges connect $A$
vertices.  In any dimer configuration in $G^*$, every $B$ vertex
must be covered by a dimer and this dimer must end at an $A$ vertex. Since the numbers of
$A$ and $B$ sites are equal, these $AB$-dimers cover all vertices.  Thus,
the inserted edges do not enter the picture. Q.E.D.

\medskip
In ensuing sections  individual lattices are considered in separate sections. 
   The Archimedean nomenclature  \cite{gs} of tilings 
are also indicated.
  
\section{The simple-quartic lattice ($4^4$)}
For a simple-quartic lattice with uniform dimer weights $z_1$ and $z_2$ in the two
(horizontal and vertical) directions, we have  \cite{kas} - \cite{fisher61}
(see also section 4 below)
\be
f_{\rm SQ}(z_1,z_2) = \frac 1 {8\pi^2} \int_0^{2\pi} d\t\int_0^{2\pi} d\phi 
\ln \big[
 2(z_1^2 + z_2^2 -z_1^2 \cos(\t+\phi) - z_2^2 \cos(\t-\phi ) \big]. \label{square} 
\ee
The free energy $f_{\rm SQ}(z_1, z_2)$ is regular in $z_1$ and $z_2$.
 
\medskip
Setting $z_1=z_2=1$ and  making use of the integration identity (\ref{id1}), we obtain
\bea
S_{\rm SQ} &=& \frac 1 {8\pi^2} \int_0^{2\pi} d\t\int_0^{2\pi} d\phi 
\ln \big[
 4(1-\cos\t\cos\phi ) \big] \nonumber \\
&=& \frac 1 {\pi} \int_0^{\pi/2} d\t \ln \big[2(1+ \sin \t)\big] \nonumber \\
&=& \frac 4 {\pi} \int_0^{\pi/4} d\t \ln (2 \cos \t) \nonumber \\
&=& \frac 2 \pi \ G  \label{SQG}
\eea
where the last step follows from the identity 
(4.224.5) of \cite{GR}  and
\bea
G = 1 -3^{-2} + 5^{-2} - 7^{-2} + \cdots = 0.915\ 965\ 594... ...\nonumber
\eea
is the Catalan constant.
It follows that we have
\bea
 S_{\rm SQ} &=& 0.583\ 121\ 808... \nonumber \\
W _{\rm SQ} &=& 1.791\ 622\ 812 ...
\label{square1}
\eea
   
\section{The honeycomb lattice ($6^3$)}
Kasteleyn \cite{kas1} has pointed out that phase transitions can occur
in dimer systems with anisotropic weights and cited the honeycomb lattice 
as an example. The honeycomb dimer problem describes
a modified KDP model  whose statistical property was
 later analyzed in details \cite{wu68, wu67}.
Here is a summary of the findings.

\medskip
Let $z_1,z_2,z_3$ be the dimer weights along the three edge directions of the
honeycomb lattice.
We have \cite{wu68} (see also section 4 below)
\bea
f_{\rm HC}(z_1,z_2,z_3) &=& \frac 1 {8\pi^2} \int_0^{2\pi} d\t\int_0^{2\pi} d\phi 
\ln \big[
 z_1 ^2 +z_2^2+z_3^2 + 2z_1z_2 \cos \t \nonumber \\
  && \hskip 2.5cm + 2z_2z_3 \cos \phi  +2z_3z_1\cos (\t-\phi).
 \big] \label{honeycomb}
\eea
When one of the 
dimer weight dominates so that $z_1, z_2, z_3$ do not form a triangle,
the free energy (\ref{honeycomb}) is 
frozen and one has
 \be
f_{\rm HC}(z_1,z_2,z_3) = \ln z_i, \hskip 1.5cm z_i \geq z_j+z_k, \hskip 1cm i,j,k {\rm \>\> distinct}.
\ee
The second derivatives of the free energy in the $\{z_1,z_2,z_3\}$-space
exhibit 
an inverse square-root singularity near the phase boundaries  $z_i = z_j+z_k$.
We shall refer to this singular behavior as the KDP-type transition.

\medskip
Setting $z_1=z_2=z_3=1$ and carrying out the $\phi$-integration,  we obtain 
\bea
S_{\rm HC}&=&   \frac 1 {8\pi^2} \int_0^{2\pi} d\t\int_0^{2\pi} d\phi 
\ln \big[3+2\cos \t + 2 \cos \phi  +2\cos (\t-\phi) \big]\nonumber \\
&=& \frac 1 {4\pi} \int_{-\pi}^{\pi} d\t \ln {1\over 2}\Big[ 
  3+ 2 \cos \t +\big|1+2\cos \t\big| \Big] \nonumber \\
&=& \frac 1 {4\pi} \int_{-2\pi/3}^{2\pi/3} d\t \ln (2+2\cos\t) \nonumber \\
&=& {2\over \pi} \int_0^{\pi/3} d \t \ln (2\cos \t ) \nonumber \\
&=& 0.323\ 065\ 947... \label{HCG} \\
W _{\rm HC} &=&  1.381\ 356\ 444 ...
 \label{hcentropy}
\eea
The resemblance of the last integral in (\ref{HCG}) with
that in (\ref{SQG}) is striking. We also remark that
the entropy $S_{HC}$ is the same as that of the ground state 
of an isotropic antiferromagnetic Ising model \cite{bh}.
The  integral in (\ref{hcentropy}) was first obtained and evaluated 
 by Wannier  in a  study of the latter problem
more than half century ago \cite{wannier}.
 
\medskip
Finally, we point out that
the honeycomb free energy can also be evaluated 
when there exists a dimer-dimer interaction \cite{huang}.  Analyses
of the phase diagram and the associated critical behavior
make use of the method of the Bethe ansatz and are fairly involved.
Readers are referred to
\cite{huang} for details.

\section{The checkerboard square lattice}
The checkerboard lattice is a simple-quartic lattice with anisotropic dimer weights
$z_1,z_2,z_3$ and $z_4$ as shown in Fig. 1.  Again, the solution of this problem
was certainly known to Kasteleyn \cite{kas1} who cited that the model exhibits 
phase transitions.  The  solution has also been mentioned by Montroll 
\cite{mont}, and studied recently in some detail by Cohn, Kenyon and Popp \cite{propp}.
 Here we provide a concise analysis using previously known results on vertex models.

\medskip
Orient  lattice edges  as shown for which
it is known  \cite{ttwu} that the Kasteleyn clockwise odd sign rule \cite{kas}
can be realized by setting $z_2 \to iz_2, \ z_4 \to iz_4$ in the 
evaluation of the Pfaffian.  
This permits us to
take unit cells of two lattice sites as shown in Fig. 1. 
Since the cells form a rectangular array, 
following the standard procedure \cite{mont} one obtains its
 free energy 
given by 
the generally valid expression
  \be
f(\{z_i\}) = 
\frac 1 {8\pi^2} \int_0^{2\pi} d\t\int_0^{2\pi} d\phi \ln \det F(\t, \phi), \label{general}
\ee
where
\bea
 F(\t, \phi)&=& M_{(0,0)} +M_{(1,0)} e^{i\t}+a(-1,0) e^{-i\t} +M_{(0,1)} e^{i\phi}
    \nonumber \\
&&  +\  M_{(0,-1)} e^{-i\phi}+M_{(1,1)} e^{i(\t+\phi)}+M_{(-1,-1)} e^{-i(\t+\phi)} ,
\eea
and the $M$'s are matrices reflecting the orientation and
connectivity of the edges. 

\medskip
For the checkerboard lattice we have\footnote{The convention used here in writing down (\ref{sqmatrix})
is such that the direction to the right in Fig. 1 is the $(1,1)$-direction.}
\bea
M_{(0,0)}&=& \begin{pmatrix}0& z_1\cr -z_1&0 \end{pmatrix}, \quad
M_{(1,0)} = \begin{pmatrix}0& 0\cr -iz_4&0 \end{pmatrix}, \quad
M_{(-1,0)} = \begin{pmatrix}0& iz_4\cr 0&0 \end{pmatrix},\nonumber \\
M_{(0,1)} &=& \begin{pmatrix}0& 0\cr iz_2& 0  \end{pmatrix}, \quad
M_{(0,-1)} = \begin{pmatrix}0& -iz_2 \cr 0& 0\end{pmatrix}, \nonumber \\
M_{(1,1)} &=& \begin{pmatrix}0& 0\cr z_3&0 \end{pmatrix},\quad\  
M_{(-1,-1)} = \begin{pmatrix}0& -z_3\cr 0&0 \end{pmatrix}. \label{sqmatrix}
\eea 
Explicitly, this leads to
 \bea
&& f_{\rm CKB}(z_1,z_2,z_3,z_4) = \frac 1 {8\pi^2} \int_0^{2\pi} d\t\int_0^{2\pi} d\phi \ln 
  \big| z_1-iz_4e^{i\t} + iz_2 e^{i\phi} -z_3 e^{i(\t+\phi)} \big|^2 \nonumber \\
  &=&  \frac 1 {8\pi^2} \int_0^{2\pi} d\t\int_0^{2\pi} d\phi \ln 
\big[z_1^2+z_2^2 +z_3^2+z_4^2 + 2(z_1z_4-z_2z_3) \sin \t  \nonumber \\
  && \hskip 1cm -  2(z_1z_2-z_3z_4) \sin \phi
-2z_1z_3 \cos(\t+\phi) -2z_2z_4 \cos(\t-\phi) \big] \ . \label{ckd}
\eea
 For $z_1=z_3, z_2=z_4$ the solution reduces 
to (\ref{square}) for the simple-quartic lattice
with uniform weights, and for $z_4=0$ the solution reduces to
(\ref{honeycomb}) for the honeycomb lattice.  Comparing the second line of (\ref{ckd}) 
(after changing $\t \to \pi/2 - \t, \ \phi \to \pi/2 - \phi$)
with Eq. (16) of \cite{fanwu}, we see that the checkerboard dimer 
model is completely equivalent to a free-fermion 6-vertex model with  weights
\bea
&& \omega_1=z_4, \quad \omega_2=z_2, \quad \omega_3=z_1, \quad \omega_4=z_3 \nonumber \\
&& \omega_5\omega_6 = 0, \quad \quad  \quad \quad \quad \omega_7\omega_8 = z_1z_3+z_2z_4\ .\label{6weights}
\eea
 
\medskip
 To analyze the free energy it is most convenient to apply
 the integration formula (\ref{id2})
to the first line of (\ref{ckd}). 
This gives
 \be
 f_{\rm CKB}(z_1,z_2,z_3,z_4) = \frac 1 {2\pi} \int_0^{2\pi} d\t \ln 
{\rm max} \big\{ |z_1+iz_4e^{-i\t}|, |\ iz_2+z_3e^{-i\t}| \big\}. \label{ckb}
\ee 
It is then seen that the free energy $f_{\rm CKB}$ is analytic in $z_i$, except when one
of the weights dominates so that $z_1,z_2,z_3,z_4$ do not form a quadrilateral, namely,
\be
2z_i \geq z_1+z_2+z_3+z_4, \hskip 1cm i=1,2,3,4. \label{ckboundary}
\ee
When this happens the free energy is frozen and 
\be
f_{\rm CKB}(z_1,z_2,z_3,z_4) = \ln z_i\ .\label{ckfreeze}
\ee

The free-fermion 6-vertex model with free energy (\ref{ckb}) has been studied in details by
Wu and Lin \cite{wulin}.\footnote{The model (\ref{6weights})  is transformed into
one discussed in \cite{wulin} after reversing 
all horizontal arrows in the vertex configurations.  As
 found in  \cite{wulin}, results reported therein  apply to uniform as well as staggered models.}
It is found that, provided that either
$z_1\neq z_3$ or $z_2\neq z_4$ (or both), namely, it is not the 
uniform  simple-quartic model discussed in section 2,
the second derivatives of $f_{\rm CKB}$
in the $\{z_1,z_2,z_3,z_4\}$-space exhibit an inverse square-root singularity
of the KDP-type transition 
near the phase boundary (\ref{ckfreeze}).  This shows that the uniform model
of section 2 is a unique
degenerate case for which the free energy is analytic.

\section{The triangular lattice ($3^6$)}
The study of dimers on the triangular lattice has been of interest
for many years (see, for example, \cite{na} - \cite{fendley}).  Although an  edge orientation of the
triangular lattice satisfying the Kasteleyn clockwise-odd sign rule
\cite{kas} for a Pfaffian evaluation has been given by Montroll \cite{mont},
it seems that the closed-form  expression of the solution appeared in print only very
recently  \cite{kenyon, fendley}.
  Here,  we provide a derivation of the solution 
using the Montroll edge orientation and analyze  its physical properties.
 
\medskip
Divide the lattice into unit cells containing two sites  as shown
in Fig. 2. 
Then, the free energy is given by the general expression (\ref{general}) with
\bea
M_{(0,0)}&=& \begin{pmatrix}0& z_1\cr -z_1&0 \end{pmatrix}, \quad
M_{(1,0)} = \begin{pmatrix}0& 0\cr z_1&0 \end{pmatrix}, \quad
M_{(-1,0)} = \begin{pmatrix}0& -z_1\cr 0&0 \end{pmatrix},\nonumber \\
M_{(0,1)} &=& \begin{pmatrix}z_2& -z_3\cr 0&-z_2  \end{pmatrix}, \quad
M_{(0,-1)} = \begin{pmatrix}-z_2& 0\cr z_3&z_2\end{pmatrix}, \nonumber \\
M_{(1,1)} &=& \begin{pmatrix}0& 0\cr z_3&0 \end{pmatrix},\hskip 1cm
M_{(-1,-1)} = \begin{pmatrix}0& -z_3\cr 0&0 \end{pmatrix}.
\eea 
This leads to
\bea
&&f_{\rm TRI}(z_1,z_2,z_3) \nonumber\\
&=& \frac 1 {8\pi^2} \int_0^{2\pi} d\t\int_0^{2\pi} d\phi \ln 2
\big[z_1^2+z_2^2+z_3^2 -z_1^2 \cos \t
 -z_2^2\cos2\phi  +z_3^2\cos(\t+2\phi)\big]\nonumber \\
&=& \frac 1 {8\pi^2} \int_0^{2\pi} d\a\int_0^{2\pi} d\b \ln 2
\big[z_1^2+z_2^2+z_3^2 +z_1^2 \cos \a 
+z_2^2\cos\b  +z_3^2\cos(\a+\b)\big], \nonumber \\
 \label{tri}
\eea
where in the last step we have  changed variables to $\a=\pi-\t,\ \b = \pi -2\phi$.
The free energy (\ref{tri}) is of the form of that of a free-fermion 8-vertex
model considered by Fan and Wu \cite{fanwu}.
Comparing (\ref{tri}) with Eq. (16) of \cite{fanwu},
one obtains the free-fermion vertex weights $\omega_1, \cdots, \omega_8$ as given by
\bea
 \omega_1^2+\omega_2^2+\omega_3^2+\omega_4^2 &=& 2(z_1^2+z_2^2+z_3^2) \nonumber\\
 \omega_1\omega_3 - \omega_2\omega_4 &=& z_1^2 \nonumber \\
 \omega_1\omega_4 - \omega_2\omega_3 &=& z_2^2 \nonumber \\
 \omega_3\omega_4 - \omega_1\omega_2 &=& z_3^2 \nonumber \\
\omega_5 \omega_6&=& \omega_1 \omega_2 \nonumber \\
\omega_7 \omega_8&=&\omega_3 \omega_4. \label{omegas}
\eea
It was found \cite{fanwu} that the model exhibits a transition at the critical point
\be
(-\omega_1+ \omega_2+\omega_3+ \omega_4)(\omega_1- \omega_2+\omega_3+ 
\omega_4)(\omega_1+ \omega_2-\omega_3+ \omega_4)
(\omega_1+ \omega_2+\omega_3- \omega_4)=0\ ,
 \label{ffcri}
\ee
and that the transition is of
 a KDP-type transition with an inverse square-root singularity in the second derivative
of the free energy if 
$\omega_1\omega_2\omega_3\omega_4 = 0$, and  an Ising-type transition 
with a logarithmic singularity in the second derivative 
if \  $\omega_1\omega_2\omega_3\omega_4 \neq 0$.

\medskip
We can solve for the $\omega$'s by
 forming linear combinations of the
equalities in (\ref{omegas}) to obtain
\bea
(-\omega_1+ \omega_2+\omega_3+ \omega_4)^2 &=& 4 z_3^2 \nonumber \\
(\omega_1- \omega_2+\omega_3+ \omega_4)^2 &=& 4( z_1^2+z_2^2+z_3^2) \nonumber \\
(\omega_1+ \omega_2-\omega_3+ \omega_4)^2 &=& 4z_2^2 \nonumber \\
(\omega_1+ \omega_2+\omega_3- \omega_4)^2 &=& 4z_1^2. \label{triweight}
\eea
This leads to the explicit solution
\bea
\omega_1 &=& \frac 1 2 \Bigg[z_1 + z_2 -z_3 + \sqrt{z_1^2+z_2^2+z_3^2}\ \Bigg] \nonumber \\
\omega_2 &=& \frac 1 2 \Bigg[z_1 + z_2 +z_3 - \sqrt{z_1^2+z_2^2+z_3^2}\ \Bigg] \nonumber \\
\omega_3 &=&  \frac 1 2 \Bigg[z_1 - z_2 +z_3 + \sqrt{z_1^2+z_2^2+z_3^2}\ \Bigg] \nonumber \\
\omega_4 &=& \frac 1 2 \Bigg[-z_1 + z_2 +z_3 + \sqrt{z_1^2+z_2^2+z_3^2}\ \Bigg] 
\eea
using which one
  verifies that we have $\omega_1\omega_2\omega_3\omega_4 \neq 0$.
This  implies that the triangular dimer model
 exhibits Ising-type transitions at 
\be
 z_i=0,\quad  i=1,2,3.
\ee
Namely, the uniform simple-quartic model discussed in section 2
  is the critical manifold of the triangular model.

\medskip
 Setting $z_1=z_2=z_3=1$,   we obtain
\bea
S_{\rm TRI} &=&
 \frac 1 {8\pi^2} \int_0^{2\pi} d\a\int_0^{2\pi} d\b \ln 
\big[6 + 2 \cos \a 
+ 2\cos\b  + 2\cos(\a+\b)\big] \nonumber \\
&=& \frac 1 {2\pi} \int_0^{\pi} d\a \ln \Big[3+\cos \a +\sqrt{7+4\cos\a +\cos^2 \a} \Big] \nonumber \\
  &=& 0.857\ 189\ 074\ ... \nonumber \\
W_{\rm TRI} &=& 2.356\ 527\ 353... \label{trientropy}
\eea
It is of interest to note  that an integral similar to (\ref{trientropy}),
\bea
S_{\rm SPT}&=&  \frac 1 {4\pi^2} \int_0^{2\pi} d\a\int_0^{2\pi} d\b 
\ln \big[6-2\cos \a - 2 \cos \b  -2\cos (\a+\b) \big] \nonumber \\
&=& \frac 1 \pi \int_0^{\pi} d\a \ln \Big[3-\cos \a +\sqrt{7-8\cos\a +\cos^2 \a} \Big] 
\nonumber \\
&=& 1.615\ 329\ 736\ ...\ ,
  \label{hcsptree}
\eea
which  gives the per-site free energy of spanning trees on the triangular lattice \cite{wu77, sw},
is reducible to a simple numerical series akin to the Catalan constant. 
 The reduction involves the mapping of the spanning tree problem to a Potts model and in turn 
to an $F$ model on the triangular lattice \cite{baxter68}.  After some steps \cite{wu033}
the spanning tree entropy (\ref{hcsptree}) becomes
\be
S_{\rm SPT} =\frac {3\sqrt 3} { \pi} \big(1 - 5^{-2} + 7^{-2} - 11^{-2} + 13^{-2} -...\big)
 = 1.615\ 329\ 736\ ...\label{hcsptree1}
\ee
 This  suggests the possibility that other entropy expressions
  can perhaps be similarly reduced.

\section{The kagom\'e lattice ($3\cdot 6\cdot 3\cdot 6$)}
The kagom\'e lattice is shown in Fig. 3 with dimer weights
$z_1,z_2,z_3$ along the 3 edge directions.
The study of  the molecular freedom for the kagom\'e lattice 
has been a subject matter of interest for many years (see, for example, \cite{pw1,elser}), but
most of the studies have been  numerical or approximate. 
    
\medskip
The kagom\'e free energy has been computed recently
\cite{wu03} by using its equivalence 
  to a staggered 8-vertex model  solved by Hsue, et. al \cite{hsue}.  
The solution is   found to be surprisingly simple and is given by
\be
f_{\rm KG}(z_1,z_2,z_3) = \frac 1{3} \ln (4z_1z_2z_3).  \label{kg}
\ee
The solution (\ref{kg}) differs fundamentally from those of
  other lattices as it does not have a series expansion. 
   This explains why most of  other approaches,
  which are invariably based on  series
expansions, are not very effective.\footnote{However, the series
expansion scheme is recovered if one introduces further symmetry-breaking 
weights \cite{wu03}.}

\medskip
  From (\ref{kg}) 
we now have
\bea
S_{\rm KG} &=& \frac 2 3 \ln 2 = 0.462\ 098\ 120... \nonumber \\
W_{\rm KG} &=& 2^{2/3} = 1.587\ 401\ 051 ...
\eea

\section{The 3-12 lattice ($3\cdot 12^2$)}
The 3-12 lattice is shown in Fig. 4.
We consider the case of six different dimer weights $x,y,z,u,v,w$.

\medskip
The 3-12  lattice has been used by Fisher \cite{fisher68} in a dimer formulation
of  the Ising model.  Using Fisher's edge orientation which we show  in Fig. 4,
  one finds the free energy given by the general expression (\ref{general}) with
\be
F(\t,\phi) = \begin{pmatrix}0 &x &z &0&-we^{-i\t}&0\cr
                            -x& 0& y&0&0& -ve^{-i\phi}\cr
                -z&-y & 0&u &0 &0 \cr
                0& 0& -u& 0& z& y \cr
                we^{i\t} &0 &0& -z& 0& x\cr
                0& we^{i\phi} & 0 & -y & -x& 0 \end{pmatrix}.
\ee
 Namely,
  \bea
 &&f_{3-12}(x,y,z;u,v,w) \nonumber \\
&=& \frac 1 3 \cdot \frac 1 {8\pi^2} \int_0^{2\pi} d\t\int_0^{2\pi} d\phi 
 \ln \big[ \omega_1^2 +\omega_2^2 +\omega_3^2 +\omega_4^2  
 +2(\omega_1\omega_3-\omega_2\omega_4)\cos\t \nonumber \\
&& \quad \quad +2(\omega_1\omega_4-\omega_2\omega_3) \cos\phi 
 +2(\omega_3\omega_4-\omega_1\omega_2)\cos(\t-\phi) \big] 
\eea
  where the factor $1/3$ is due to the fact that there are 3 dimers per unit cell, and
\be
\omega_1=x^2u,\quad \omega_2=uvw,\quad \omega_3=y^2w, \quad \omega_4 =z^2v\ . \label{3-12}
\ee

The free energy (\ref{3-12}) is  again of the form of that of a free-fermion 8-vertex model
 with vertex weights $\omega_1, \omega_2, \omega_3, \omega_4$,   $\omega_5\omega_6 = \omega_3\omega_4$, and
$ \omega_7\omega_8=\omega_1\omega_2$ and the critical point 
 (\ref{ffcri}).  For the Ising model 
with interactions $K_1$ and $K_2$   \cite{fisher68}, for example, we have
$x=y=z=u=1$, $v=\tanh K_1$, $w=\tanh K_2$.  It is then verified that 
the critical condition
 $\omega_1=\omega_2+\omega_3+\omega_4$ can be realized and gives the known Ising critical point
\be
\sinh 2K_1 \sinh 2K_2 =1.
\ee
Finally, setting $x=y=z=u=v=w=1$, we obtain
\bea
S_{3-12}&=& \frac 1 3 \ln 2  = 0.231\ 049\ 060... \nonumber \\
W_{3-12} &=& 2^{1/3} = 1.259\ 921\ 049...
\eea

\section{The 4-8 lattice ($4\cdot 8^2$)} 
Dimer models on the  4-8 lattice
  shown in Fig. 5 have been used to describe  phase transitions in physical
systems \cite{allen} -  \cite{nagle}.  
 By setting $u=v$, for example, the dimer model describes the phase transition in the
layered hydrogen-bonded  $SnCl^2\cdot 2H_2O$ crystal \cite{sn}. 

\medskip
Orient the lattice as shown, and this leads to  the free energy  
(\ref{general}) with
\be
F(\t,\phi) = \begin{pmatrix} 0&xe^{i\t}&-v&-u\cr
                             -xe^{-i\t}& 0& u& -v \cr
					v& -u & 0& ye^{i\phi} \cr
					u&v&-ye^{-i\phi} & 0 \end{pmatrix}. \label{4-8}
\ee
Namely,
  \bea
 f_{4-8}(x,y;u,v) 
&=& \frac 1 2 \cdot \frac 1 {8\pi^2} \int_0^{2\pi} d\t\int_0^{2\pi} d\phi 
 \ln \Big[ x^2y^2 +(u^2+v^2)^2 \nonumber \\
&& \quad  -2xyu^2\cos (\t+\phi) -2 xy v^2 \cos(\t-\phi) \Big] .
 \eea
 
\medskip
The free energy (\ref{4-8}) is also of the form of the that of a free-fermion 
8-vertex model \cite{fanwu}
and exhibits an Ising type transition at 
\be
u^2+v^2 = xy.
\ee

\medskip
Setting $x=y=u=v=1$, one obtains
\bea
S_{4-8} &=& \frac 1{16\pi^2} \int_0^{2\pi} d\t\int_0^{2\pi} d\phi 
 \ln\Big[5-2\cos(\t+\phi) -2\cos (\t-\phi)\Big] \nonumber \\
&=&  \frac 1{16\pi^2} \int_0^{2\pi} d\t\int_0^{2\pi} d\phi 
 \ln \big[5-4\cos\t \cos \phi \big] \nonumber \\
&=&  \frac 1{2\pi} \int_0^{\pi/2} d\t \ln \Bigg[ \frac {5+\sqrt{25-16\cos^2\t}}2
  \Bigg] \nonumber \\
&=& 0.376\ 995\ 650\ ... \nonumber \\
W_{4-8} &=& 1.457\ 897\ 968\ ...
\eea
  
\section{The Union Jack lattice [$4\cdot 8^2$]}
The Union Jack lattice shown in Fig. 6(a) is the dual of the 4-8 lattice.  
It is constructed
by inserting diagonal edges with weights $u$ and $v$ to a checkerboard lattice. 
Since the checkerboard lattice is bipartite and the inserted diagonals  connect vertices of
one sublattice only, 
  the Proposition established in section 1 now implies the identity
\be
Z_{\rm UJ} (z_1,z_2,z_3,z_4;u,v) = Z_{\rm CKB} (z_1,z_2,z_3,z_4) .
 \ee
That is, the solution for the Union Jack lattice is identical to 
that of a simple-quartic lattice as if the $u$, $v$ edges were absent.  
Particularly, they have identical
entropy and molecular freedom.

\section{The [$3\cdot 12^2$] tiling lattice}
The [$3\cdot 12^2$] tiling lattice shown in Fig. 6(b) is the dual of the 3-12 lattice.
It consists of two sets of vertices,  set $A$ vertices each of which having coordination number 3 
and set $B$ 
vertices each having coordination number 12.  The number of $A$ vertices is twice that of $B$,
and the $A$ vertices are connected to $B$ vertices only.

\medskip
The lattice [$3\cdot 12^2$] does not admit dimer coverings.  This follows from the fact
that in a proper dimer covering each $A$ vertex must by covered by a dimer and the dimer
must end at a $B$ vertex. This means some $B$ vertices will have  more than one dimers 
and this is not possible.  Thus, there is no proper dimer
coverings and the generating function $Z_{[3\cdot 12^2]}$  is identically zero. 

 \section{Summary and Acknowledgments}
We have presented   analytic and numerical results on the free energy, entropy, and molecular
freedom for close-packed dimers on two-dimensional lattices which have
unit cells arranged on a rectangular array.  
  For the anisotropic checkerboard lattice the free energy is found to exhibit a KDP-type singularity,
  except in the degenerate case of
uniform dimer weights the free energy is analytic.  
 For the triangular lattice  the free energy is analytic for nonzero $z_i$ 
and is critical at $z_i=0$.   For  4-8 and 3-12 lattices the
dimer models exhibit Ising-type transitions. 
  
\medskip
Numerical results on the entropy and molecular freedom are summarized in Table 1.
We observe  that the entropy is not necessarily a monotonic 
function of the coordination number of the lattice.  
Analyses of other Archimedean tiling lattices 
still remains open.

\medskip
I would like to thank R. Shrock for  introducing me to 
Archimedean tilings and for comments and a critical reading of the manuscript,
J. Propp for calling my attention to Ref. \cite{propp}, and F. Hucht for pointing
a numerical error in Eq. (\ref{hcsptree1}) in an earlier version of the manuscript.
 The assistance of W. T. Lu in preparing the figures is also gratefully acknowledged.
Work has been supported in part by NSF Grant 
  DMR-9980440.
 
\begin{table}

\begin{tabular}{|l|l|l|l|} \hline 
Lattice   & \quad Entropy $S$ &\ Molecular freedom  &Phase \\
& & \quad \quad $W=e^S$ & transitions \\ \hline \hline
Honeycomb ($6^3$)   & \quad 0.323\ 065\ 947  & \quad 1.381\ 356\ 444&\quad KDP \\
Checkerboard  ($4^4$)           & \quad 0.583\ 121\ 808 & \quad 1.791\ 622\ 812 &\quad KDP\\
Triangular ($3^6$)  & \quad 0.857\ 189\ 074 & \quad  2.356\ 527\ 353 &\quad  Ising\\
Kagom\'e ($3\cdot 6\cdot 3\cdot 6$) & \quad 0.462\ 098\ 120 & \quad  1.587\ 401\ 051 &\quad None\\
3-12 ($3\cdot 12^2$)  & \quad 0.231\ 049\ 060 & \quad 1.259\ 921\ 049&\quad Ising\\
4-8 ($4\cdot 8^2$)  & \quad 0.376\ 995\ 650 &\quad  1.457\ 897\ 968&\quad Ising\\
Union Jack [$4\cdot 8^2$]  & \quad 0.583\ 121\ 808 & \quad 1.791\ 622\ 812&\quad Ising \\
$[3\cdot 12^2]$ & \quad no dimer coverings & \quad  no dimer coverings &\quad None\\
\hline
  \end{tabular}
\caption{Summary of numerical results. Phase transitions occur for
anisotropic dimer weights}

\end{table} 
  
 \hbox{}

\newpage

\centerline{Figure Captions}
\vskip 1cm

\bigskip
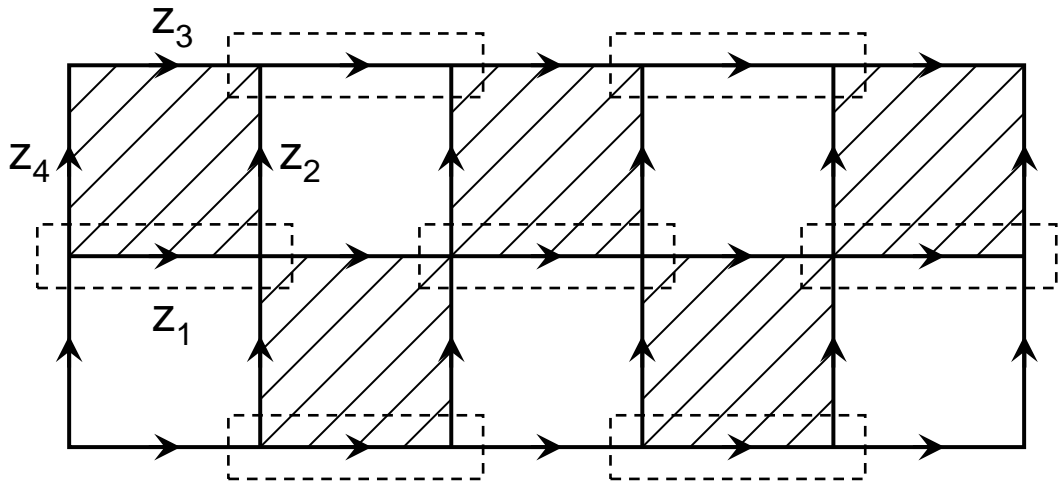
Fig. 1. Edge orientation and unit cells of the checkerboard lattice.  Shaded squares
are repeated.  Weights $z_2$ and $z_4$ are replaced by $iz_2$ and $iz_4$
inn the evaluation of the Pfaffian (see text).

\bigskip
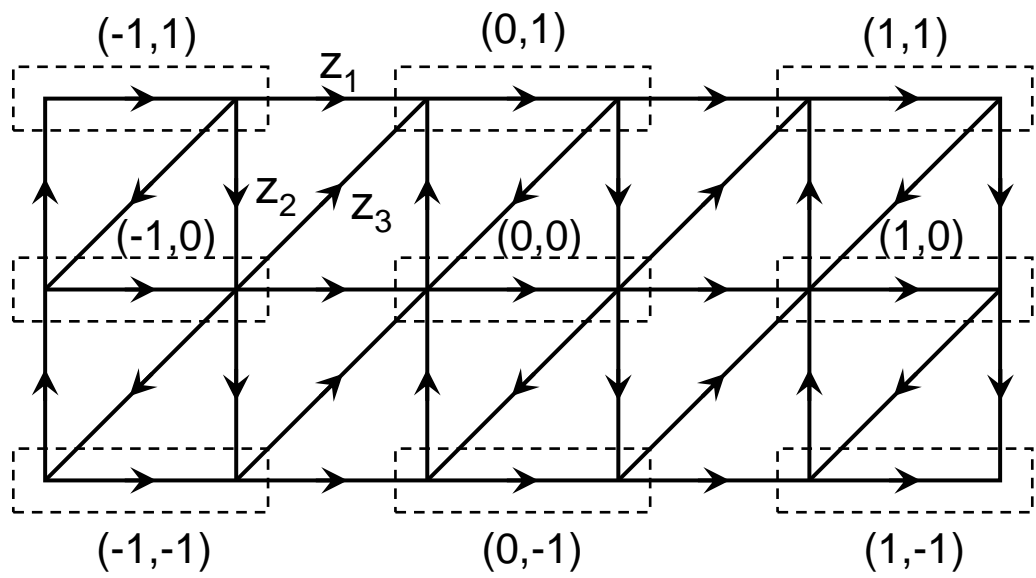
 Fig. 2. Edge orientation and unit cells of the triangular lattice.

\bigskip
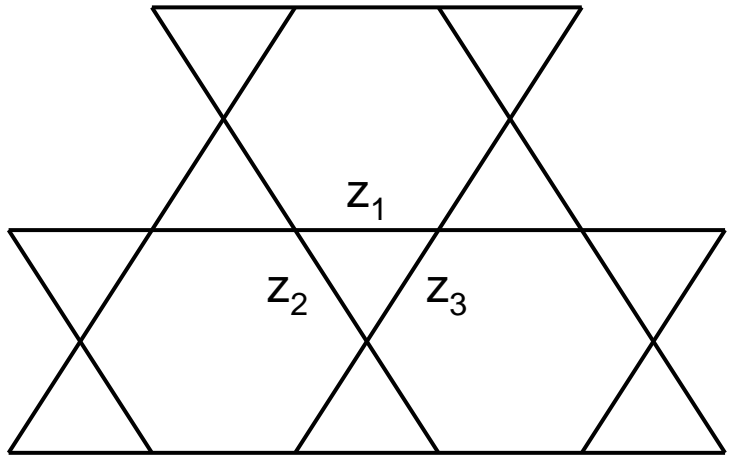
Fig. 3. The kagom\'e lattice.

\bigskip
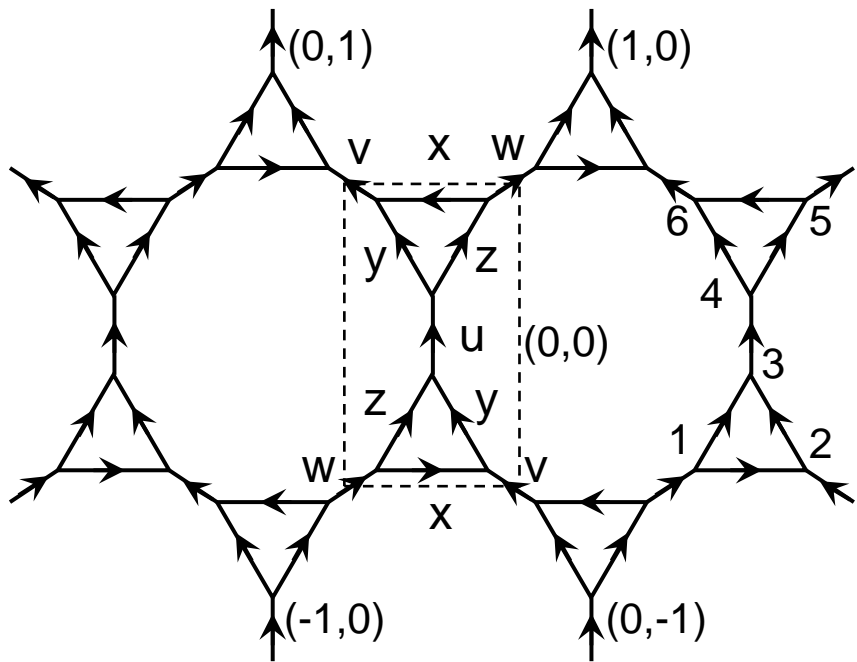
Fig. 4. Edge orientation and a unit cell of the 3-12  lattice.

\bigskip
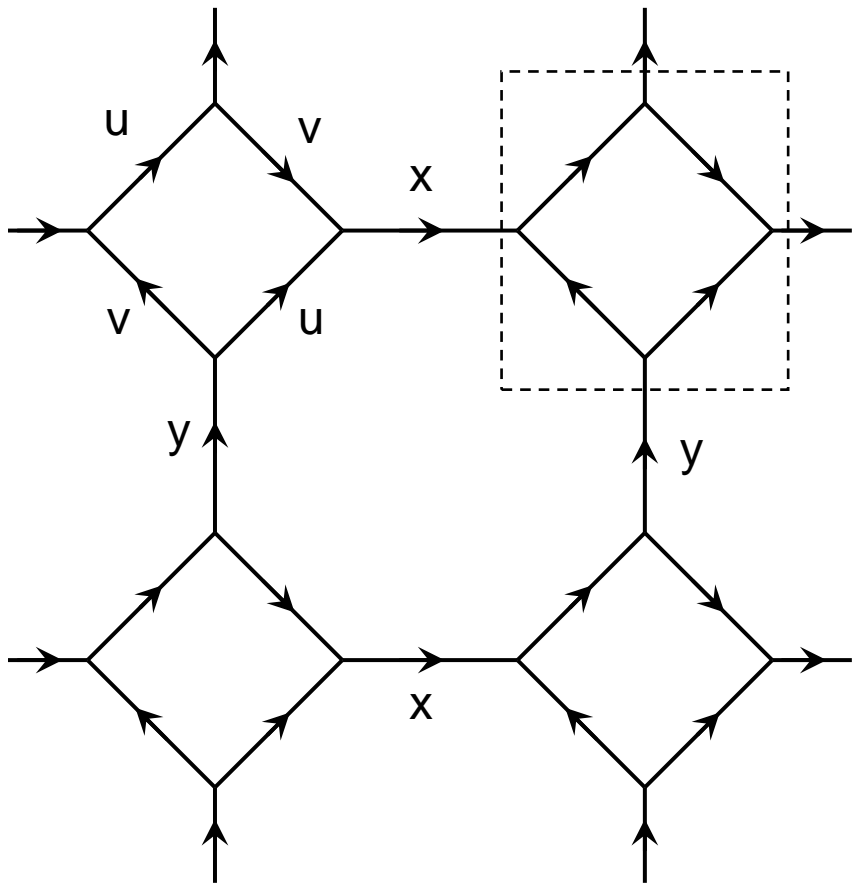
Fig. 5. Edge orientation and a unit cell of the 4-8 lattice.

\bigskip
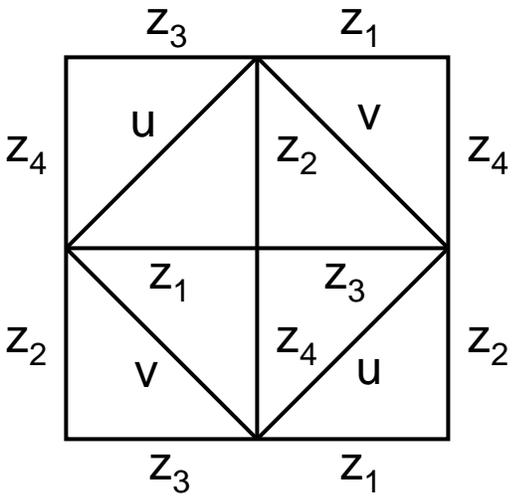
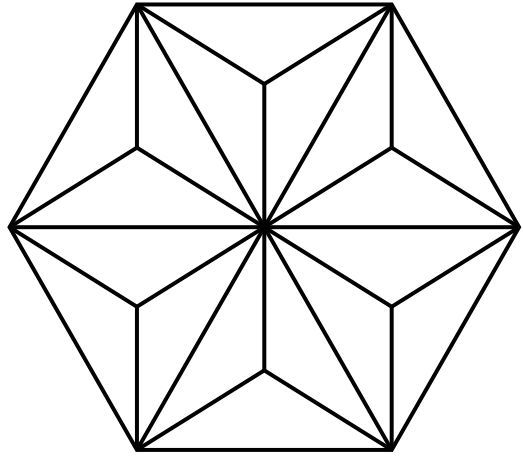
Fig. 6. (a) The [$4\cdot 8^2$] Union Jack lattice, the dual of the 4-8 lattice.
         (b) The [$3\cdot {12}^2$] lattice, the dual of the 3-12 lattice.
\end{document}